# On the natural selection of macromolecular network and biological complexity

Jacques H. Daniel (1) and (2)

((1) Centre National de la Recherche Scientifique, Centre de Génétique Moléculaire, Gif-sur-Yvette, France, (2) InvenTsion, Rehovot, Israel, as present address;  Correspondence: jacques.daniel1@gmail.com)

**Modern biological tools have made it possible to unequivocally demonstrate the deep relationship among species in terms of genes and basic molecular mechanisms. In addition, results from genetic, physical and physiological approaches applied to individual model cell systems have led to the mapping of large networks of macromolecular interactions with similar general properties. Although gene, mechanism and network structure similarities among species tend to suggest the existence of important constraints applied to organisms, it is surprising that the elucidation of the precise general mechanisms by which natural selection could have operated appears to have taken a back seat in most, if not all, studies concerning the evolutionary development of cell macromolecular networks. Herein, a possible explanation is presented for how cells could have evolved into the sophisticated and integrated organisms we know today, using natural selection. Based on the new concept of gene toxicity, essentially applied to unicellular organisms with their rapid multiplication rate, it is proposed that natural selection likely exerts its effects in two opposing and complementary directions  to integrate a new gene function into the host system; a simple and basic qualitative model for the buildup of cell macromolecular networks is suggested. Finally, as a radical consequence, it is proposed that complex life evolution has proceeded through a long alternation of unicellular and multicellular states.**

**Introduction**

The complexity of the living world is astonishing in terms of the number and diversity of organisms, and variations in organization. One obvious fracture in this world is the profound divide between unicellular and multicellular organisms, although the makeup of both is extremely intricate and efficient, marking their long persistence and adaptation in the course of evolution.

Unicellular organisms appear unique in their capacity to reproduce exponentially and at a very high rate, given the proper conditions. Indeed, their very small size makes them ideal substrates for the incremental and rapid construction of ever new living units, at a relatively low energetic cost (tiny volume and large surface to volume ratio). Although --and also, because-- these cells are quite complex, they can make very fast copies of themselves that are totally identical, except for some occasional small changes. This powerful and rapid "circularity", coupled to the capacity for random genetic variations (such as gene duplication, lateral gene transfer, mutation), is a unique feature of unicellular organisms and as such may have great relevance to our explanation of the mechanisms at the origin of biological evolution.

It is clear that unicellular organisms preceded the appearance of multicellular organisms, which were derived from them, with a higher level of complexity. However, both types of biological organization were never totally insulated in the course of evolution, in particular being able to make even more complex physical and functional associations called holobionts, in both in the plant and animal kingdoms (Rosenberg and Zilber-Rosenberg 2016).

Present-day life is essentially based on the functioning of macromolecular networks that are generally extremely complex, implicating numerous types of activities and a formidable number of interactions among the protein, DNA and RNA species involved. As a rule, these networks rely on positive, as well as negative, functional interactions, generally giving rise to the smooth adaptive and balanced behavior characteristic of every living system. In addition to their balanced growth, some systems have the ability to embark into differentiation as a result of the unrolling of strict built-in genetic programs.

Obviously, these macromolecular networks, with their specific characteristics (i.e., the precise types of functional interactions involved) could not have arisen in the absence of selection. However, defining the particular selective forces underlying the buildup of this macromolecular complexity during evolution remains a difficult challenge. As a proposition for helping to solve this issue, I wish to point to some general observations made on the behavior of genes in a unicellular model organism.

**About genes: ambiguity and constraints**

Genes, being at the very essence of life, are rightly assumed to be beneficial components in the functioning of cells and organisms. This is obvious for genes whose disruption leads to inviability but it also holds true for the numerous so-called dispensable genes because they have persisted in the recipient organisms throughout eons of evolution. However, genes do not act alone: being part of a system, they are likely to be subject to constraints as a result of their integration into this wider assembly. Indeed, some basic considerations of gene products strongly suggest this view to be correct.

According to the selective theory of allostery, many proteins exist in (at least) two conformational states that are in equilibrium, only one of them being the active form. Thus, any physiological regulation of protein activity implies a shift from one conformational form to the other, i.e., to the active (activation) or inactive (inhibition) form. This situation places a rather strong constraint on the cell's concentration of proteins because too much of a particular protein, for instance, might result, by the simple equilibrium between its forms, in the production a constant activity signal instead of the normal modulation for maximum physiological efficiency. This concentration constraint is even more salient if strict stochiometry between a regulatory and catalytic subunit is required to form a complex: an instance of this is provided by the yeast protein kinase A whose normally inactive protein complex breaks down when cAMP binds to its regulatory subunit, thus freeing protein kinase A activity. This point is not merely theoretical: it has been shown that overexpressing the catalytic subunit of protein kinase A overcomes the growth arrest in cells harboring a deficient activator of adenylate cyclase which is normally responsible for cAMP synthesis, and is thus one essential intermediate in the signal-transduction pathway involving this nucleotide (Daniel and Simchen 1986). Indeed, there are innumerable instances of similar situations, which have led to the recognition of suppression by extragenic overexpression as a powerful genetic approach for the study of macromolecular networks in yeast.

Thus, one could ask whether in a normal cell bearing *no detrimental mutations*, an over-dosage of any gene might have some adverse effect on the cell's economy, possibly by preventing the occurrence of the right activity at the appropriate time, for example, during the cell cycle, or in response to growth conditions or various stresses.

A very sensitive and convenient method was developed in yeast to test the effect of gene over-dosage on cell fitness. It involves using a specific yeast recipient strain that contains a multicopy plasmid bearing the gene of interest, and looking at the rate of plasmid loss (that is, by not selecting for the plasmid markers) in the growing cell population, as compared to the loss rate of the control vector (i.e., in the absence of the tested gene): the more rapid the apparent loss rate, the more toxic the gene maintained at high dosage (Daniel 1993, 1996b). Strikingly, as presented by some examples in Table 1, all of the genes tested to date under these conditions have indeed been found to exhibit some toxicity, ranging from low to very high values (see footnote of Table 1 for more experimental details).

This general phenomenon made it possible to devise a novel genetic approach (called fitness-based interferential genetics or FIG) to elucidate the functional interactions occurring in vivo within the cell's macromolecular networks (Daniel 1993, 1996a, 2005, 2007, 2009). In a nutshell, one selects, in a genomic library made on a similar compatible multicopy vector (with different specific markers), for genes that modify the toxicity (i.e., the plasmid loss rate) of a particular gene (the bait) borne on the first multicopy vector, meaning that the two genes functionally interact.

The observed toxicity that is taken advantage of here appears to be primarily due to overexpression of the specific gene --although a high dosage of its native promoter region may have a minor contribution -- as strongly suggested by the following results: (i) the possibility of neutralizing protein kinase A toxicity by concomitant high cAMP binding protein gene dosage (Daniel 1996a) (ii) the selection of a bona fide protein target by FIG involving the protein kinase A gene (Daniel 1993) (iii) the selection of the unique known inhibitor of protein kinase C by FIG involving a highly toxic mutated form of protein kinase C (Daniel 2009) (iv) a few other FIG selections in which the nature of the obtained genes clearly implies overexpression of a negative transcription factor as the bait, as well as their own overexpression (Daniel 2005, 2007, 2009) (for more arguments along these lines see below in Table 3, Fig. 1, and text).

In addition to its potentially extensive use for functional interactomics, and of primary concern for the present work, the implication of this general observation made on genes for evolution seems considerable: genes appear to have both beneficial and "toxic" effects. This apparent paradox can be demonstrated for dispensable genes using the following simple setup: one measures the toxic effect of a gene in a wild-type recipient cell as afore described, and compares its value to that obtained in an isogenic strain that is disrupted in its chromosomal gene counterpart (Table 2). One can observe the gene toxicity phenomenon, reflected by a more rapid plasmid loss rate as compared to the control vector, in the wild-type strain; however, in the disrupted strains, plasmid loss is significantly decreased relative to the wild type, and in one case, nearly reversed. Faced with a dilemma, the disrupted strains therefore tend to keep the plasmid whose gene corresponds to the one lacking on their chromosome (indicating that this gene is indeed beneficial), even though its overexpression is clearly toxic. Breslow et al. (2008) arrived at the same conclusion about the general beneficial effect of genes by directly measuring the fitness of strains in a yeast deletion library.

As to the general gene toxicity phenomenon, even more remarkable results were found when different heterologous genes, of various origins, were expressed under a yeast constitutive promoter borne on the above yeast plasmid. As shown in Table 3, all of the genes lead to toxicity in yeast, many quite markedly, with one (human *BRCA1*) even being the most toxic gene ever measured under any condition. Moreover, no correlation was observed between toxicity and the length of the overexpressed foreign protein, meaning that the toxicity phenomenon is likely linked to some specific function of the particular protein involved (Fig. 1). For practical purposes, this toxicity of heterologous genes has been used for the screening of small inhibitory molecules targeted against their specific protein products (Daniel 2008), or for detecting functional interactions between these proteins and potentially active small ligands (Castello et al. 2011; unpublished). Interestingly, from an evolutionary perspective, such experiments using heterologous genes may somehow resemble the situation of a cell facing the advent of new genes during evolution, with their eventual possible benefit as well as their toxicity.

Herein, to devise any theory to account for the increase in biological complexity during evolution, it is contended that these two aspects of a gene --utility *and* toxicity-- should be borne in mind; one possible simple and general scenario is suggested.

**The tripartite model for increasing macromolecular network complexity**

The Janus face of genes thus affords the possibility during evolution of selecting not only *for* new gene adoption, but also *against* gene "excess" following this adoption. In other words, it provides a means for allowing the integration of a useful new gene within the recipient cell system.

The following basic model is proposed to account for gene acquisition. As a first step, some new function X emerges via modification of a duplicated gene (or, possibly, a foreign gene introduced by lateral genetic diffusion). At some point, a coupling Y/Z is made from an *already existing* gene product Z and a new (or already somehow existing, see below) gene product Y rendered inactive by Z's activity. Finally, Y becomes an inhibitor of X, with the result that X is active only when Y is inactivated by Z. Thus, Z positively controls X as a result of a sequence of two inhibitions (Fig. 2).

In this process, selection occurs on two (mainly successive) fronts. One concerns the new function X, which adds an interesting feature to the cell (this is, of course, in relation to the beneficial side of a gene). The other front involves the (indirect) control of X by Z, which allows for the integration of X within the cell system (this is in relation to the "no excess" requirement for a gene, to be considered in broad terms, that is, correct quantity and correct availability relative to timing within the cell cycle, the growth environment, specific stress conditions, and so on).

The term "inhibition" by Z or Y is understood to have a broad meaning: it can occur either at the level of gene expression (repression) or against protein activity (inactivation). Moreover, direct positive control of X by Z is not favored because it would require X to be inactive by default. This is unlikely since X's activity has to be selected for in the first place, X being a metabolic enzyme, a component of a macromolecular complex, a signal-transduction pathway, etc. Y serves as a messenger, linking Z to X, the link between Y and Z first occurring as happenstance, and then selection fixing a Y that binds and inhibits X unless it is modified by Z (again, both events can occur at either the gene or protein level). It should be stressed that the proposed model for increasing biological complexity represents only a basic scenario: obviously, the final picture of the networks is likely to result from many refinements fixed later on as a result of mutual adaptation of the constitutive components.

Thus, it is proposed that the hard core of this control system is made up of an intermediate Y, acting negatively on the new gene X, and itself regulated negatively by an existing gene Z that happens to be "physiologically close" to the X function. Since several other functions (i.e., X', X"…) might be related to that of X (in terms of the timing of their requirement during the cell cycle, or the requirements of the milieu and so on), it is likely that some of these functions might also be controlled by Z, following a mechanism similar to that postulated for X. This may contribute --at least in part-- to the power law distribution of functional macromolecular interactions generally observed in cellular networks (Albert 2005). Indeed, the presumed pre-existence of several different Ys as intermediates in the control of these existing functions by Z might make it easier to create additional Ys, since it increases the likelihood of obtaining random duplications of these genes ("the rich getting richer"). Once a Y gene has been duplicated, its modification (to Y') could occur --spontaneously or even by selection against some gene-dosage toxicity effect-- in a region of possible interaction with a new X (i.e., X', again as a gene or protein), leaving the site of interaction with Z unchanged (indeed, loss of interaction with Z together with the established interaction with X' would likely be unfavorable by exerting a dominant negative effect on X'). These events will thus place X' under Z's control.

Another element might contribute to the power law distribution of functional macromolecular interactions (instead of an exponential law, for example) might be related to some specific types of molecular control having been preferentially selected for being more versatile, simple and efficient (for instance, protein phosphorylation vs. other types of interaction).

Notably, this model is compatible with the recent findings of a bigger than expected number of repressors in the yeast genome (Kemmeren et al. 2014). Moreover, this model might account for two salient features which appear to characterize cells' macromolecular networks, i.e., hierarchy and modularity (von Mering et al. 2003; Barabasi and Oltvai 2004; Goodman and Feldman 2017). Indeed, the proposed Z activity could control several components (U, V, W…) other than X, through Y or other inhibiting (or even activating) intermediate components. As a result, a hierarchically controlled module could be formed, which would be even larger if some master controlling element regulating Z existed upstream, or if some "growth" occurred downstream by some of the components (U, V, W, X…) acquiring regulatory functions in turn.

It should be stressed, with regard to present-day cell components, that there is no easy way to assign their primitive status as being of either the X or Y type, since the considerable growth in cell complexity that has occurred during evolution is expected to have completely blurred this information. Thus, for instance, a present-day protein kinase might have initially emerged as an X-type function, improving the efficiency of some cellular process, or alternatively, it might have started as a Y-type intermediate for inhibiting some particular target. In agreement with this, Barabasi and Oltvai (2004) found it difficult to precisely delineate modules, and modules of modules, within various macromolecular networks.

The overall aspect of the cell's macromolecular network might be foreseen as follows. An essential and specific feature of this network could be its grossly *cyclical* structure, primarily as a reflection of cell-cycle dynamics (indeed, one of the most primitive characteristics of expansion-prone living systems such as unicellular organisms), even though other smaller sub cycles may well be fitted in. In fact, this property may already be apparent in one rough sketch of the yeast protein network (Fig. 3a): it seems that a hollow is indeed present at its center, whereas a computer-generated scale-free power-law network based on a gene duplication/mutation model does not appear to display such a feature (Fig. 3b). In this yeast protein network, emerging branches are appended to the crucial control elements making up the circular (and evolving) backbone of this network, growing into domains and modules with evolutionary time (Fig. 3a).

**The progressive evolution to multicellularity**

Whatever the mechanism actually used for increasing macromolecular network complexity, it seems inescapable, after the invention of a new biological function, that a consolidation stage be added, aimed at integrating this novelty into the general cell machinery. The unicellular state, with its exponential and relatively rapid growth, appears to be the most efficiently poised for performing such an accomplishment by cell competition and natural selection. However, the most sophisticated biological forms are certainly to be found among multicellular organisms, thus posing a major challenge to the evolution scientist in determining how these very complex biological systems could ever have arisen and developed.

Here, one possible general scenario is proposed for the creation and expansion of the multicellular life form: it is based on alternation, during evolution, between unicellular and multicellular (monoclonal) states. This could ensure benefiting from both the unicellular world --for efficiently integrating any new valuable function into the existing macromolecular network--, and the multicellular world --for inventing all types of biological architectures and devices adapted to various environments.

One striking --and perhaps somewhat overlooked-- feature of cells from any multicellular system is their relatively rapid growth, which allows rapid building-up of the organism and its efficient maintenance. Indeed, the speed of forming a multicellular organism might be critical for taking full advantage of a new environment or outcompeting other organisms; and in this context, having rapid repair capacity (i.e., replacing deficient cells) might also be an invaluable asset. Therefore, the unicellular states, in between the multicellular ones, may accomplish the necessary "smoothing" of the cell's macromolecular network following any favorable genetic change that had been tested previously. Later in evolution, this perpetual smoothing and persistent fitness adjustment might indeed have been critical for the appearance and success of germ lineages as well as immune systems.

It should be stressed that the unicellular life form has additional advantages: (i) it can more readily absorb new genes by lateral diffusion, which may help the building-up of newer or better multicellular systems, or the gene-integration process; (ii) it can undergo gene shuffling and enrichment by sexual means. Thus, all of the benefits brought out by the unicellular state would have amplified effects on the multicellularity state progression, up to the stage of consolidation and quasi exclusivity, reached by the existing multicellular species populating the living world that we normally observe.

One difficult issue for evolution studies concerns the mechanisms involved in the very primordial stage of cell differentiation giving rise to the very first pluricellular organisms. Although the jury is still out with respect to understanding these precise driving mechanisms, one general avenue might, nevertheless, be suggested: the cell's organization into functionally consistent subsets under hierarchical control may have made it easier to differentially express these various subsets, thus leading to differentiation into the various tissues characteristic of pluricellular systems. As some crude examples of this, one might speculate on how cells could have overdeveloped the protein subset essentially involved in the function of vacuoles so as to lead to the creation of a primitive exocrine pancreas organ; on how they could have overexpressed subsets involved in membrane and cytoskeleton growth and secretion systems, thus resulting in some primitive neurons; or on how they could have "fused" by inhibiting the division and cytokinesis subsets and, by overexpressing the subset related to the development of motor fibers, given rise to primitive muscle cells, and so on. Remarkably, a family of dinoflagellates (Warnowiaceae) displays complex eye spots ("ocelloids") in their cytoplasm that are very similar to camera-type eyes, with cornea-, lens-, and pigment cup-like elements, changing structure in response to changing illumination, and apparently containing rhodopsin related to bacterial rhodopsin (Hayakawa et al. 2015). Presumably, the specific switch that allows cell differentiation by overexpressing one functional subset in certain cells, and not others, would have originally come from the differential exposure of the cells to a particular environment. Later in the evolution of multicellular organisms, this switch would have resulted from the unfolding of an internal program of development caused by some drastic dosage imbalances in gene regulation (Birchler et al. 2005; Conrad and Antonarakis 2007) and/or duplicated conserved noncoding elements (McEwen et al. 2006).

A decisive question is how the long alternation between unicellular and multicellular states contemplated here could have been made possible and effective. Firstly, multicellularity --as we can study it-- has occurred at least 25 times during evolution in all natural kingdoms, thus suggesting that this might not be such an exceptionally rare event (Parfrey and Lahr 2013; Niklas 2014). Furthermore, in the bacterial kingdom, there are signs that multicellularity might have already occurred three and a half billion years ago (Westall et al. 2006). Secondly, it has been theoretically shown that multicellularity might be a quite trivial outcome, appearing "as an inevitable consequence of dynamical systems" (Furusawa and Kaneko 2002). Finally, a serious and fundamental issue should be addressed on how this unicellular/multicellular "community of fate" could have operated in practice during evolution. Although this is obviously a rather difficult question, it might be suggested -- considering the "combinatorial mapping "of unicellular macromolecular network modules unto the multicellular differentiated cells, as proposed above--that any genetic event (i.e., a new gene, or a mutation/modification of an existing gene, for example) could have been tested for being a valid improvement, in both the unicellular and multicellular states, and if it happened to be so, could eventually have undergone the integration/regulatory process in the unicellular state, as discussed above.

**Concluding remarks**

We are used to thinking of the advent of multicellularity as a rather brisk and definitive transition from unicellular life, made possible by the acquisition of certain genes responsible for cell-cell attachment or extracellular matrix formation, for instance. The reality, however, may be less clear-cut, involving, as proposed herein, a long evolutionary period of back and forths between multicellular and unicellular states, i.e., some kind of long-lasting multi-sequential cooperation. The main reason for that would have been the amplification loop constituted by the multicellularity's strength in inventing potentially unlimited variations of biological organizations and apparatuses adapted to the environmental and ecological context, coupled to the fitness-oriented regulatory rigor characterized by the unicellular state, thus leading to a formidable win-win situation marking the history of life on Earth (Fig. 4). Although this hypothesis --essentially suggested by the complex gene behavior found in yeast cells-- needs to be further tested, analysis of the genome of two unicellular organisms -- a choanoflagellate, and a filasterean--seems to be compatible with this view because, as one group of authors stated in their paper's title (Suga et al. 2013), the genome revealed "a complex unicellular prehistory of animals" (King et al. 2008; Brunet and King 2017). It is expected that results of many more future genomic and physiological studies, and maybe evolutionary experiments in vitro, will permit the testing of the hypothesis presented here of a profound and prolonged collaboration between unicellular and multicellular organisms for the building-up of complex macromolecular networks and the increase in biological complexity during natural evolution.

**Figure Legends**

**Figure 1.** Absence of correlation between length of foreign proteins and toxicity. Protein length is given in numbers of constitutive amino acids (AA). $R^2$ equals 0.0813 for linear regression and 0.0837 for exponential regression. See Table 3.

**Figure 2**. The tripartite model for increasing biological complexity. For details, see text.

**Figure 3**. (a) Topology of a real yeast proteome map. (b) Topology of a typical computer simulation of the network model obtained by gene duplication (Pastor-Satorras et al. 2003) (both maps extracted by permission). For details, see text.

**Figure 4**. The back and forth (or accelerator) model of biological complexity increase during evolution.

**Table 1**: Toxicity of gene overexpression

___________________________________________________________________

| Gene | TPK1 | SIK1 | RPL4A | BCY1 | ROG1 | CTF18 | VID27 | INO4 | SSL1 |
|---|---|---|---|---|---|---|---|---|---|
| **Toxicity*** | 62 | 31 | 25 | 16 | 13 | 12 | 11 | 11 | 10 |

===================================================================

| Gene | RVS167 | MID2 | STE13 | CWC21 | MLP1 | RPD3 | RAS2 | PKC1 | SUR1 |
|---|---|---|---|---|---|---|---|---|---|
| **Toxicity*** | 10 | 9 | 9 | 7 | 6 | 3 | 2 | 2 | 1 |

___________________________________________________________________

*Recipient yeast strain was C90-A (*leu2 ade2Δ*). Genes with their own promoter regions were cloned into multicopy vector YEp21-bA (containing both the *LEU2* and *ADE2* markers), and plasmid loss rate in cultures in the presence of leucine and moderate concentrations of adenine was monitored thanks to the red-pigment-accumulating property of cells regaining their adenine requirement, by using the three-streak test (Daniel 1993, 1996b). Numbers represent the differences in the percentage of white colonies out of total (white + red) colonies between cultures carrying the vector and cultures carrying the various gene plasmids: the higher this value, the more toxic the gene. Between 150 and 300 colonies were screened for each gene. For rationale, see text.

**Table 2**: Unmasking the beneficial side of dispensable genes

| Gene | Toxicity* in wild-type cells | Toxicity* in disrupted cells | Reduction of toxicity in disrupted *vs.* wild-type cells | |
|---|---|---|---|---|
| | (A) | (B) | (A-B) | (A/B) |
| *CTF18* | 17.8 | 3.6 | 14.2 | 5 |
| *MID2* | 19.9 | 0.3 | 19.6 | 66 |
| *RPL4A* | 44.5 | 28.3 | 16.2 | 1.6 |
| *VID27* | 18.0 | 8.1 | 9.9 | 2.2 |

*Yeast strains were *ade2Δ* derivatives (this work) of BY4741 (wild type) and of the derived gene-disrupted strains (obtained from Euroscarf). See footnote in Table1.

**Table 3**: Toxicity of foreign genes expressed in yeast

| **Gene** | Luciferase | Integrase | *BRCA1* | *DBP1* | *DIP2* | Hexokinase1 | Hexokinase2 |
|---|---|---|---|---|---|---|---|
| **Origin** | *Firefly* | *HIV* | *Human* | *Arabidopsis* | *Arabidopsis* | *Arabidopsis* | *Rat* |
| **Toxicity*** | 10 | 5 | 66 | 10 | 18 | 24 | 20 |

*See footnote in Table 1 for strain and experimental procedure. Foreign genes of diverse origins were placed under the control of the yeast *PGK1* constitutive promoter carried on a derivative of the YEp21-bA multicopy vector.

**Figure 1:**

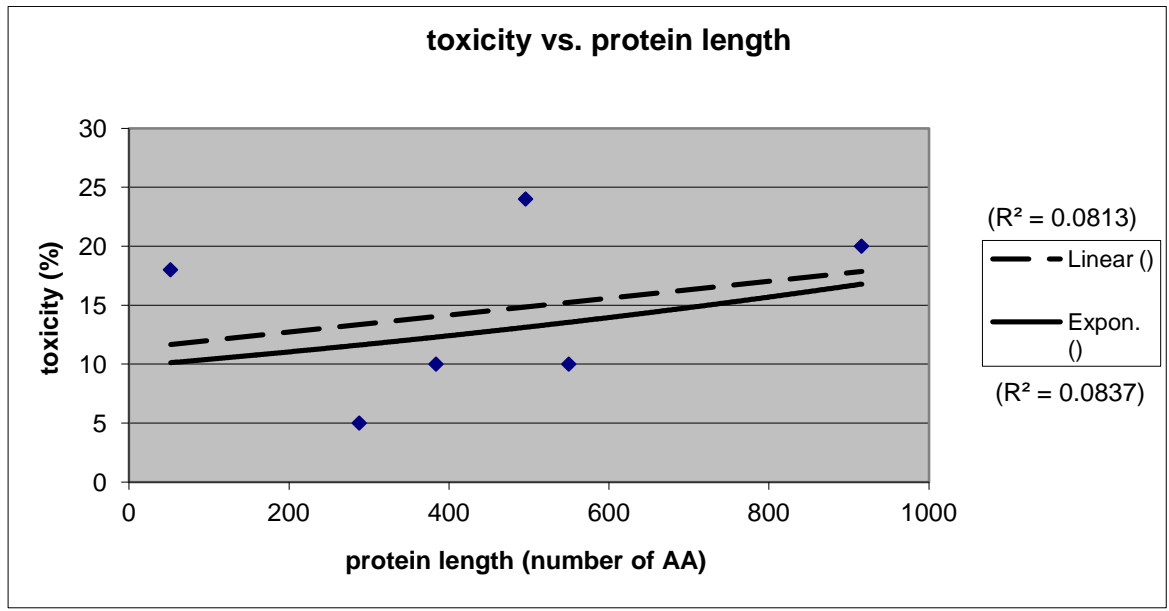

**Figure 2:**

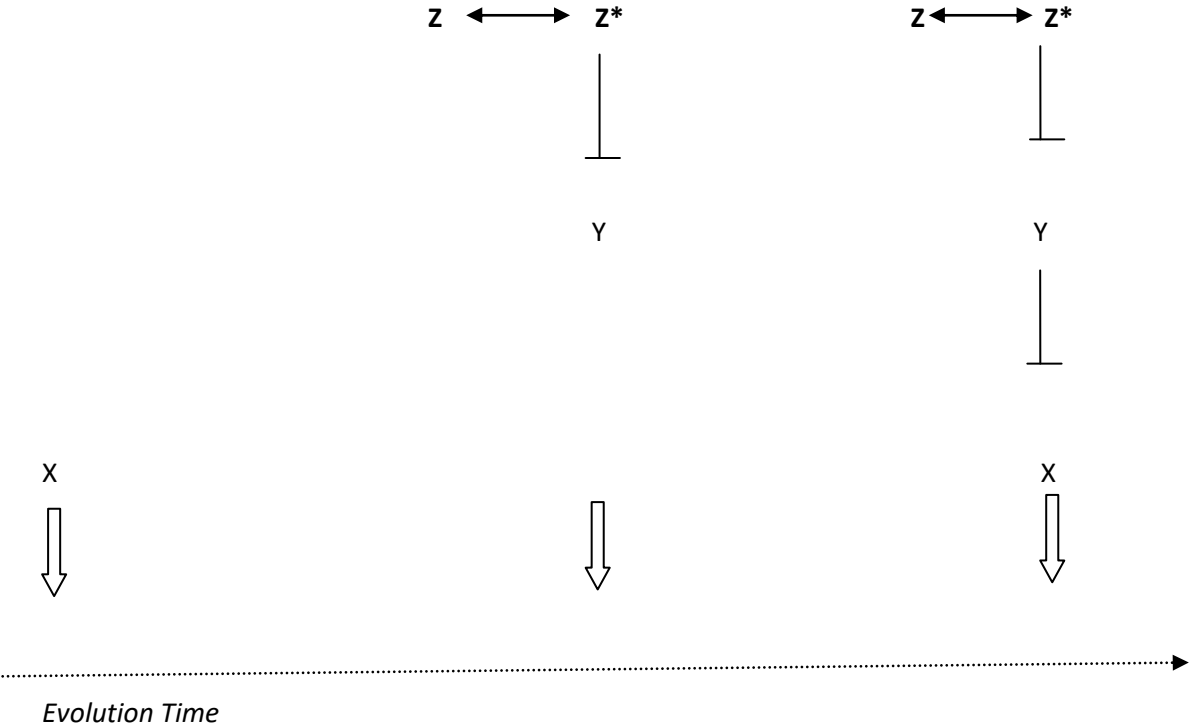

*Evolution Time*

**Figure 3**:

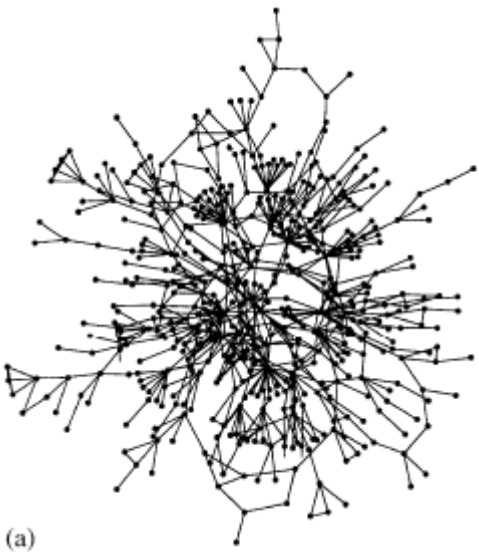 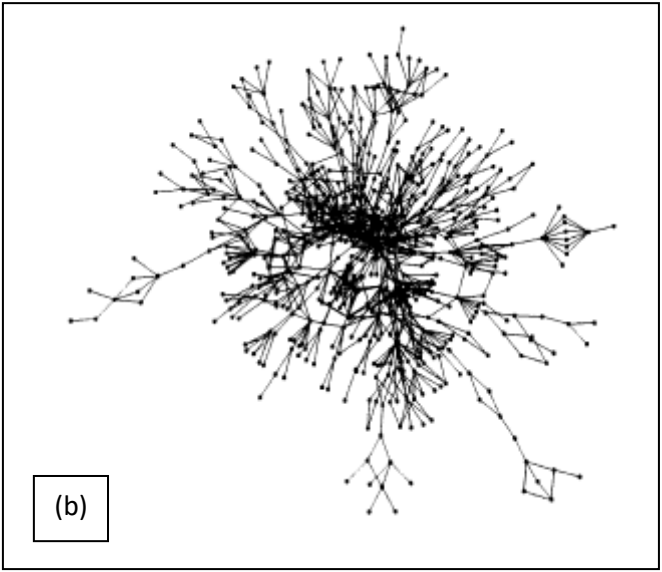

**Figure 4:**

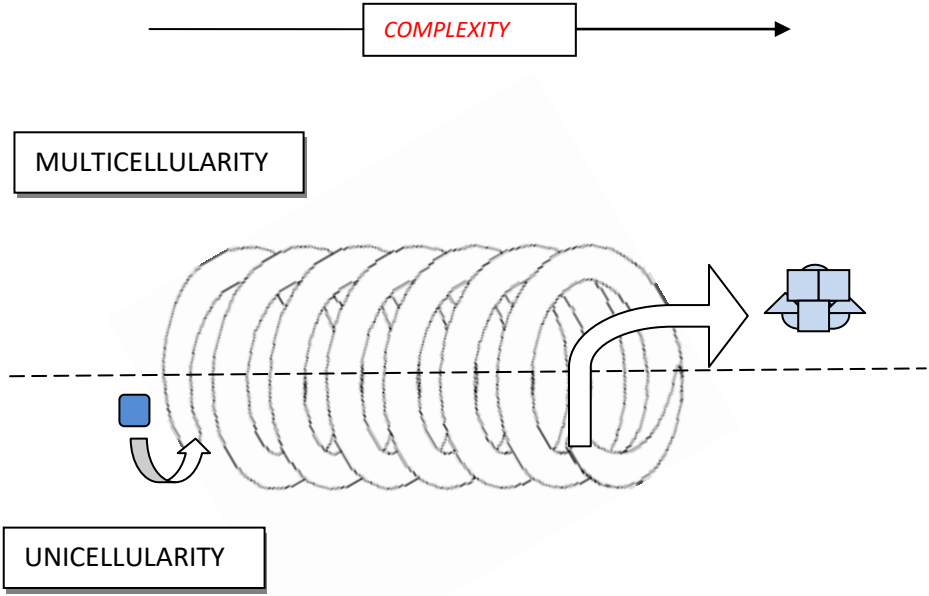